# Directed Testing of ORAN using a Partially Specified Declarative Digital Twin

Alan Gatherer
Cirrus360
Dallas, USA
gatherer@cirrus3sixty.com

Chaitali Sengupta
Cirrus360
Dallas, USA
chaitali@cirrus3sixty.com

Sudipta Sen
Cirrus360
Dallas, USA
sudipta@cirrus3sixty.com

Jeffery H. Reed
Cirrus360 and Virginia Tech
Blacksburg, USA
reedjh@cirrus3sixty.com

*Abstract*—**Real Time performance testing can be divided into two distinct parts: system test and algorithm test. System test checks that the right functions operate on the right data within power, latency, and other constraints under all conditions. Major RAN OEMs, put as much effort into system test and debug as they do into algorithm test, to ensure a competitive product. An algorithm tester will provide little insight into real time and hardware-software (HW-SW) capacity as it is unaware of the system implementation. In this paper we present an innovative Digital Twin technology, which we call Declarative Digital Twin (DDT). A DDT can describe the system requirements of the RAN such that critical corner cases can be found via automation, that would normally be missed by conventional testing. This is possible even when the RAN requirements are only partially specified. We present a Domain Specific Language (DSL) for declarative description of the RAN and show results from an automated solver that demonstrate how potential HW-SW implementation related corner cases can be identified from the DDT of an ORAN DU.**

## I. INTRODUCTION

Real Time performance testing can be divided into two distinct parts: system test and algorithm test. Algorithm test has been the traditional focus of test equipment manufacturers because they are testing a black box against an algorithm's performance specification with respect to metrics such as bit error rate, and probability of call drops. But at major RAN OEMs, at least as much effort goes into system test and debug of the complete RAN. This is because a competitive RAN is a finely tuned real time system operating near its capacity and any mistiming of the hardware or software can lead to unexpected functional errors. Either the RAN will crash, or its performance will degrade. But an algorithmperformance tester will provide little insight either way into what is wrong or how to fix it.

System performance testing in RAN is the act of checking for possible errors due to incorrect operation of the hardware and software in a real time environment. Common system errors include read before write, or write before read of data, timing errors such that data misses its window of correctness, buffer overflows where data is lost, and so on. Such errors tend to be occasional and may appear to happen randomly, that is as "Heisenbugs" [1], errors occurring due to specific paths through the RAN's state space. Hardware techniques such as cache coherency or locks do not solve such problems because the timing and order of access to data is function dependent and the cache/lock mechanism is unaware of the functional goals. Such errors are generally missed by traditional "highly loaded" testing methods in lab testing and trials, as explained below, but show up, often intermittently, in the field. They can also provide trap doors into the system that can be exploited by hackers.

Heisenbugs are an issue for production RAN deployments, especially in larger cellular deployments, because the RAN is a High Availability (HA) system [2]. In traditional proprietary RAN implementations, the RAN engineering team will have detailed knowledge about the hardware and software they are using and will staff a large team to focus specifically on HA functional testing. This allows them to push the envelope of capacity of the RAN, running it at high utilization. With Open RAN, the software and hardware components come from multiple vendors and are integrated and tested by potentially yet another vendor, so it becomes much harder to achieve HA while at the same time pushing towards maximum capacity.

If algorithm performance testing is applied to the RAN, the occasional system error may appear as a slight bit error rate or call drop degradation, due to a packet being dropped if data is corrupted, for instance. This may not register as a system performance loss in the test. It is also highly likely that the test suite designed for algorithm performance testing does not challenge the RAN in a way that causes any system errors. Either way the system error will slip through the testing stage and be released into the field. In the field, the RAN will experience a much more thorough set of state space paths than it experienced as part of the performance test in the lab. As a result the RAN may experience a serious error, not simply algorithmic performance degradation. These errors are usually hard to repeat in the lab as they are related to a specific path through hardware/ software state space, and may appear in the field, seemingly at random, that is, as Heisenbugs. This problem is exacerbated by the use of COTS server hardware which naturally displays some runtime uncertainty due to cache, multi-threading, out of order execution, and occasional operating system (OS) calls. All of these technologies were introduced to improve the average performance of the server in a server farm environment but will introduce seemingly random variation in runtime and data access time. If other threads are running on the system, such as edge applications, then this issue becomes more impactful [10].

A Digital Twin (DT) is generally a simulation of a system that mimics its behavior. We extend the DT concept by making

This work supported in part by NTIA PWSCIF under award 48-60-IF006

it a declarative specification. In this case the DT is not a simulation, but a declarative description of all aspects (5G/6G protocol flows, hardware/software resources, deployment scenarios and constraints), from which facts about the performance of the system can be deduced. In this paper we propose a Partially Specified Declarative Digital Twin (DDT) technology for system testing that allows the test engineer to add the details they know about the RAN functionality and implementation. This information then drives a directed test vector generation strategy, using an automated solver platform we have developed, designed to find test vectors that cause system errors, preventing these errors from being released into the field and causing problems in production.

In Section II we summarize the requirements for a Domain Specific Language [5] for RAN description. In Section III we outline the syntax of the DSL that we use to describe RAN functionality in the Digital Twin at an abstract and disaggregated level. In Section IV we describe how the automated solver platform shown in Fig. 1, called GabrielForTest™, can be used to find tests that are directed towards finding system errors in a specific hardware-software implementation. In Section V we give an example using GabrielForTest™ to evaluate the potential impact of a test scenario, for an ORAN Distributed Unit (DU), using Intel® FlexRAN™ Reference Architecture.

GabrielForTest™ generates directed test scenarios by analyzing a DDT of the device under test as well as specifications of the deployment scenario. GabrielForTest™ provides both interesting test scenarios for a tester to configure and run as well as insights into what is expected from the test. The scenarios will generally be sampled from a requirements specification. We also show a feedback loop in Fig. 1 to highlight the potential to learn declarative constraints on the behavior of the platform by observing its behavior under test, using these to enhance the DDT model.

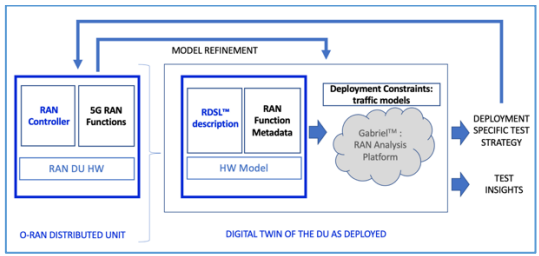

Fig. 1. GabrielForTest™ using RDSL™: How it works

## II. RAN DOMAIN SPECIFIC LANGUAGE

Prior work on automation of RAN construction prompted us to develop a DSL for RAN deployment, RDSL™. RDSL™ driven intelligent automation was recently recognized with a Global Mobile (GLOMO) award at Mobile World Congress 2024 [3]. We describe the properties of this language in this section with an emphasis on its value to test vector generation.

### A. Declarative description of the flow of data constituting the system requirements of the RAN

In a declarative language the programmer describes what must be done and not how to do it. Some popular examples include SQL, XML and YAML. The opposite of this approach is an imperative language, like C or python [7]. RDSL™ declares flows of processing in a periodic real time system, augmented by YAML [8] based constraint definitions to allow easy integration into a O-RAN DevOps flow. In YAML we include a high level description of the hardware platform. The description of the system is therefore completely declarative. The partitioning of constraints and requirements in the RDSL™ flow is shown in Fig. 2. This description defines what we know to be true about the RAN system; a Declarative Digital Twin. It is not a simulation digital twin and requires an intelligent automation platform like GabrielForTest™ to interpret it. But due to its declarative nature, new constraints and requirements can simply be added. And it does not need to contain the full set of requirements of the system, only what is known by the tester, and hence is a Partially Specified DDT. If something is undefined, then an analysis tool, in this case GabrielForTest™, is free to assume a setting for this undefined constraint that leads to a worst/best case scenario, or a set of points in between, depending on what makes sense for test vector generation.

It is common for DSLs to be declarative because they are simpler and more application focused than general purpose languages [11]. It is therefore easier to bring the benefits of declarative language to them. Some examples of declarative DSL and DSL-based frameworks include P4 for SDN and TensorFlow for ML. For configurations in data centers, there are multiple declarative languages in use including Docker, Kubernetes, Ansible and Puppet. Declarative languages are the wave of the future for safe, open, and reliable software at scale [7] and for code development within an efficient DevOps loop.

### B. An Immutable Language to expose optimization opportunities for deployment specific construction/testing

An immutable language is one in which there are no variables, only defined or undefined data elements. The lack of change in data value exposes parallelism and timing opportunities because GabrielForTest™ does not have to consider which version (value) of a data element is in use. For the purposes of test vector generation, GabrielForTest™ may assume timing that leads to errors in the system, unless there is a constraint to prevent it,

Haskell, though not a DSL, is the most famous immutable language in use today and Erlang is an immutable language used in telecoms. For DSLs, immutability is an even easier choice of property, as the language is simple and focused, and hence the lack of intuitiveness in general purpose languages such as Haskell is not present in a DSL.

## III. THE SYNTAX OF A RAN DOMAIN SPECIFIC LANGUAGE

We first define a top level view of the DSL, separated into independent entities. This is shown conceptually in Fig. 2.

### A. RDSL™ Flow syntax

RDSL™ is used to describe *flows* in the system. Each flow is a description of a group of dependent RAN functions (*modifiers*) that operate on *streams*, as shown in Fig. 3. RDSL™ has only one data type that is called a *stream*, an immutable list of infinite length that has an implicit ordering of definition so that the, conceptually, $n^{th}$ slot of the list will be defined using data from slots relative to the $n^{th}$ slot. Unless constrained, $n^{th}$ slot processing does not have to occur in the $n^{th}$ slot in real time. For instance, if processing is constrained to occur within 2 slots due to latency constraints, then GabrielForTest™ may look for test vectors where the timing of processing within that 2 slot window can be arranged to cause memory overflow, cache misses or delay due to lack of processing resources.

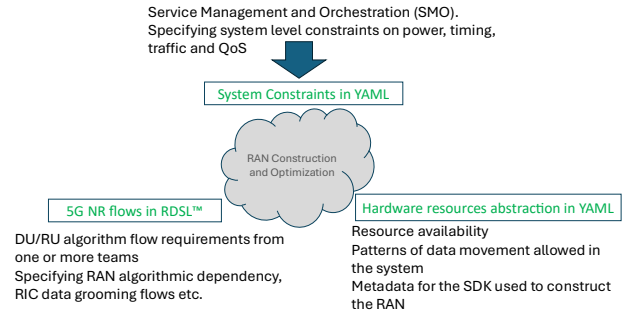

Fig. 2. Top Level Partitioning of the RAN DSL

Fig. 3. RDSL™ Syntax for a Flow

The flow in Fig. 3 calls other flows using an indexing syntax, which is a convenient way of producing multiple flows from a tensor stream. This is not an implementation but a declaration. How and when these streams are finally implemented remains flexible unless other constraints come into play. Apart from the usual "ease of use" language features, such as includes, nesting of flows (example shown in Fig. 3), local stream definitions within flows, and so on, RDSL™ also supports stream labels. These allow you to tag a stream with a unique identifier that can be used to apply timing and memory use constraints to the stream in addition to the constraints implied by the RDSL™ flow definition. For instance you can constrain a stream connected to a physical port based on the timing of that port. The constraint of this type is defined at a higher level in a Kubernetes style declaration in YAML which we describe in III.B.

### B. System Constraint Declarations

System constraints are defined individually in a YAML format that closely resembles Kubernetes [4], so that we can reuse tools such as Ansible, and because vRAN DevOps uses Kubernetes extensively and is well understood in the O-RAN ecosystem. Examples of system constraints is shown in Fig. 4.

Each constraint is defined individually, grouped into files using the "---" separator. The *apiVersion* and *kind* fields allow the parser to interpret the specification correctly. At the time of writing there are two basic *kind* formats, one which creates a simple value relationship to a constant (the left example in the figure) and one which creates a more complex relationship using an equation and multiple values. Values are symbolic and can be inherited from a network/SMO level YAML file or can be a flow label (see section III.A). They can even come from a RAN Intelligent Controller (RIC) app. In this way we can take constraints from the RIC/SMO and apply them hierarchically to the RAN design. For examples of the flexibility in applying values, the reader is encouraged to look at Ansible syntax. In Fig. 4 *grid_period* is the name of a timing value. But it may be the name of a label of a timing event or stream, or may itself be defined in terms of other labels inherited from elsewhere.

Fig. 4. System Constraint Syntax Example

Fig. 5. Hardware Resource Syntax Example

### C. Hardware Resources Constraint Declaration

Hardware resources are declared in terms of the constraints they imply on the performance of the system. Resource elements (i.e. processors, memories, interconnects, accelerators) are defined and the concept of *patterns* is introduced. A pattern is an allowed data movement of a buffer of data through the hardware from *definition* (a function writes its output to a buffer) to *observation* (a function reads the data in the buffer). We show an example in Fig. 5 in XML format. Each pattern has a name and is defined in terms of anchor memories for definition and observation. This description was generated from a higher level topological map of the hardware, including the pattern names, which were generated for human readability. But once the hardware description has been generated, it can be hand edited to restrict pattern use to align it to the middleware used by the hardware to access the resources, In the case of Intel FlexRAN™, this is the eBBUPool scheduler [9]. This is sufficient to describe potential data movement for the purposes of identifying corner cases and good test vectors that can stress the implementation.

### D. Functional Block Declaration and Metadata

Each function called in a flow has metadata describing the patterns it can use, and additional constraints on runtime and local memory usage. An example of this is shown in Fig. 6.

## IV. Finding high risk Scenarios Using RDSL™ driven intelligent automation

Given the declared constraints on the operation of the DU, GabrielForRU™ adds potential runtime failure constraints. The most interesting way to do this is to manage the patterns allowed for each stream. For instance, we can use pattern management to force data to cache out or to force a task to run on a specific subset of the cores. In this paper we also add some simple declarations that force a task to start running within a specified time after all of its input data is ready, which enforces a generic opportunistic scheduling strategy on the system [6], as the Intel FlexRAN™ uses an opportunistic scheduling strategy [9]. We then search for the *best case* runtime under all of these artificial constraints, to see if they will force a runtime failure even under otherwise good conditions. Searching for best case runtime is a strategy we employ in our constructive solver, Gabriel™ so we can reuse this code. But in GabrielForTest™ we search for failure rather than success. The analysis problem for finding good test scenarios is in fact the dual of the construction problem. Treating the DDT as a database of constraints, Gabriel™ asks the database if there exists a feasible solution to the database constraints, and if so requests a scheduling and mapping strategy within this feasible space. GabrielForTest™ adds more implementation constraints to the DDT database, including some that force the system to fail. It then asks if there is no feasible solution and if so, the DDT presents a good example of a directed test scenario.

Fig. 7 shows conceptually how this is done. The test requirements are the added constraints in the database and the model of the system under test declares the constraints that define how the platform will work for any set of requirements. Each of these requirements is a constraint set and has a feasibility region (the red and green dotted regions). GabrielForTest™ looks for solutions that occur near the edge of the overlap of these two regions, where the test requirements are still satisfied but the system is beginning to fail. Exactly how to search this space is an ongoing research topic but we show some preliminary examples in Section V.

## V. Test Vector Generation with GabrieForTest™

In this section we show an example use of RDSL™ and GabrielForTest™ to generate cache sensitivity test cases for a virtual DU (vDU) implementation. As vDU generally share a server platform (even with CPU core isolation), with other tasks and threads, such as the ORAN CU, Core, or AI/ML based edge applications cache performance is hard to control. Today's

ORAN Testing approaches have no way to design and define tests, especially in an automated manner, that specifically tests for caching behavior.

Analysis of a particular declared testcase can immediately lead to GabrielForTest™ finding that there is no feasible solution for this test on the declared platform (so no overlap of the red and green regions in Fig. 7). In this case the testcase is an obviously good choice to run in the lab on the actual DU, and optimize the code until it passes, as shown in Fig. 1. For a more interesting example of the use of GabrielForTest™ we instead pick a feasible test (overlap in Fig. 7) and add new declarations to see what the GabrielForTest™ analysis tells us about the risk faced by the testcase under these new conditions.

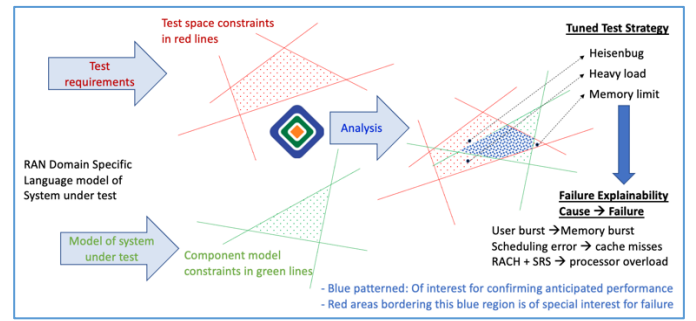

Fig. 6.   Functional block metadata  Example

Fig. 7.   From RAN Domain Specific Requirements to a Tuned Test Strategy

We consider a simple DU downlink only case, with two 10MHz downlink flows running on a 2GHz Intel Icelake Xeon Gold 6338. RDSL™ was written to reflect the tasks in FlexRAN™, and the hardware resources constraints reflect the Xeon server. Table 1 summarizes the results:

|     | Strategy | Latency (clock cycles) |
| --- | --- | --- |
| 1 | No added constraints | 207,800 |
| 2 | Small cache eviction | 239,400 (15% increase) |
| 3a | DL Config eviction | 420,000 (102% increase) |
| 3b | Large cache eviction | 464,600 (124% increase) |
| 4 | 2 + 3b | 578,000 (178% increase) |
| 5 | 3 DL no added constraints | 458,400 (120% increase) |

*Table 1: Results for different pattern management constraints*

GabrielForTest™ analysis shows that without added constraints the test can successfully complete on a hardware slice of 4 cores in about $1/5^{th}$ of the available 0.5ms slot, as shown in Case 1 of Table 1. We configured GabrielForTest™ to find the best possible latency for this test, given the constraints declared for the system. We then demonstrate the value of GabrielForTest™ analysis by adding cache sensitivity constraints to the declaration and demonstrate the impact they have upon the latency of the testcase, and therefore its usefulness for testing purposes.

First we declare that cache eviction occurs for a selection of small buffers (all under 10KB). Conceptually this moves in one of the edges of the green space in Fig. 7 allowing us to explore the space at the edge of the red/green overlap. We would not expect a large impact on the cache itself, but it may change the ordering of tasks in the opportunistic scheduler. The impact on latency was calculated by GabrielForTest™ to be about 15%, implying that the deployment is robust to cache effects for these tasks. This result is shown in Table 1 Case 2. The allowed latency constraint is an edge of the red space and this 15% increase may move the red and green spaces apart entirely.

Eviction of large buffers can lead to significant delay across the DDR interface as well as memory congestion at that interface, conceptually a large movement of an edge in Fig. 7. To test this scenario we declare the Downlink Config task evicts its data at some point (this is not unlikely as the data is widely dispersed in a large structure). Significantly, this led to a 102% increase as shown in Table 1. Case 3a. So the testcase is sensitive to the way the cache handles this data and this may in turn be sensitive to the operating system, the CaaS system, and any other tasks running on the platform, even if cores are not shared, for instance an AI task running in the background. Therefore this test may be recommended for these deployment conditions.

Interestingly, when we additionally declare that NR5G1_DL_PDSCH_TB, and NR5GG1_DL_BEAM_GEN all evict their output, the result only worsens by an additional 22%, as shown in Case 3b in Table 1. This is probably due to the dispersal of these tasks in time so that they are not quite additive. As this result is produced by analysis of a declarative twin there is the opportunity to also provide an explanation for the behavior that is human understandble. These results can explain runtime variation and predict its potential size on platforms with variable cache effect patterns.

Declaring that both the small and large buffers evict increases the latency to 578,00 clock cycles, another 54%. So the addition of the small cache eviction increases the latency by more than the increase due to the small buffer evictions alone.

As a sanity check we added a third DL flow to the system and the load increased by 120%. So the loss due to cache effects is close to the loss of a whole DL flow. So the analysis shows that cases 2-5 are all candidates for lab testing and debugging, especially in configuraions sensitive to cache behavior.

## VI. CONCLUSIONS

Using RDSL™ to declaratively describe a digital twin of the DU functionality allows for the DU to be analyzed even when there is incomplete information about the implementation and operation of the DU. Declarative programming allows new constraints and features to be easily added when discovered, and allows the test strategy to play with different conditions by declaring different possible failure conditions to see what impact they might have on the DU. These potential failure conditions then become candidates for test and debug. GabrielForTest™ can run on relatively cheap cloud servers to provide a test strategy with maximum value before the RAN is taken to the test lab and therefore there is significant potential to reduce the cost of test as well as improve its impact.

We have only scratched the surface of the potential value of RDSL™ and GabrielForTest™ for generating directed test vectors and analyzing weaknesses in DU implementations. Several new declarative failure constraint strategies are currently being developed. The potential for using AI to direct the search is clear. However, even at this early stage we can see the ability to weed out uninteresting test scenarios in favor of those most likely to produce errors, and hence requiring attention from the development team, so Heisenbugs in the field can be avoided.

## ACKNOWLEDGMENT

We would like to thank Intel[1] and Vodafone for their unwavering support and advice in the development of our automated approach to RAN deployment, TIP for their support of the early development of RDSL™, and the NTIA and USDA for their continued support of our efforts.

## REFERENCES

[1] M. Musuvathi, et al, "Finding and reproducing Heisenbugs in concurrent programs": OSDI'08: Proceedings of the 8th USENIX conference on Operating systems design and implementation, 2008.

[2] https://www.cisco.com/c/en/us/solutions/hybrid-work/what-is-high-availability.html

[3] Mobile World Congress, Barcelona., GLOMO Winner, Cirrus360 for Best Digital Tech Breakthrough for companies with under $10 million Annual Global Revenue 28 February 2024.

[4] https://dzone.com/refcardz/getting-started-kubernetes

[5] Golden Age of Computing. Hennessey and Patterson Turing Award presentation. 2017, ACM.

[6] A survey of hard real-time scheduling for multiprocessor systems. Robert I. Davis and Alan Burns. ACM Computing Surveys (CSUR), Volume 43, Issue 4 Article No.: 35, Pages 1 - 44

[7] Forge: generating a high performance DSL implementation from a declarative specification. A. K. Sujeeth, et al, ACM SIGPLAN Notices, Volume 49, Issue 3. Pages 145 - 154

[8] YAML:https://www.redhat.com/en/topics/automation/what-is-yaml . Yet Another Markup Language.

[9] 5G Cloud RAN on Intel Architecture. Intel Corp. CSAE 2023, October 17–19, 2023, Virtual Event, China

[10] G. Garcia-Aviles, A. Garcia-Saavedra, M. Gramaglia, X. Costa-Pérez, P. Serrano, A. Banchs,. (2021). Nuberu: reliable RAN virtualization in shared platforms. 749-761. ACM. 10.1145/3447993.

[11] K. Olukotun, Taming Heterogeneous Parallelism with Domain Specific Languages, DIMACS Workshop on Parallelism: A 2020 Vision, March 14-16, 2020